# Role of Carrier Mobility and Band Alignment Engineering on the Efficiency of Colloidal Quantum Dot Solar Cells

Roha Saad, and Nauman Z. Butt*, Member IEEE

Department of Electrical Engineering, School of Science and Engineering,

Lahore University of Management Science, Lahore, Pakistan

*Email: nauman.butt@lums.edu.pk

**ABSTRACT**

**We investigate physics based design of colloidal quantum dot (CQD) solar cells using self-consistent computational modeling. The significance of band alignment engineering and optimized carrier mobility are quantitatively explored as a function of sub bandgap defect densities ($N_t$) in the bulk CQD. For $N_t \leq 10^{15}$ cm$^{-3}$, band alignment engineering near the interface of CQD and the metal contact could significantly improve open circuit voltage by suppressing the forward bias dark current. This effect could enhance cell efficiency up to ~37% for thinner (<1 μm) CQD layers. For thicker (>1 μm) CQD layer, the effect of band engineering is diminished as the forward bias dark current becomes diffusion-limited and less dependent on the interfacial band offsets. An optimal carrier mobility in CQD lies in the range ~ $10^{-2}$ cm$^2$/Vs – $10^0$ cm$^2$/V-s and shows variation as a function of CQD layer thickness and the interfacial band offset. For $N_t \approx 10^{14}$ cm$^{-3}$, an optimally designed cell could provide ˜20% efficiency under AM1.5G solar spectrum without employing advanced structural optimizations such as the nanostructured electrodes. These physical insights contribute to a better understanding of quantum dot solar cell design, allowing a step further towards a highly efficient and a low cost solar cell technology.**

*Keywords*—quantum dot solar cell, carrier mobility, band alignment, efficiency

I. INTRODUCTION

Photovoltaic technology provides one of the most rapidly growing forms of renewable energy generation [1]. The growth of the photovoltaic installations around the globe is strongly linked with the cost of the solar panels [2]. The market dominant crystalline solar cell technologies [3] have a high cost of materials and expensive processes of fabrication due to which many low cost emerging materials [4-6] are being investigated for next generation solar cells. Among these, colloidal quantum dot solar cell (QDSC) is one of the attractive options for next generation solar cells [6-9]. Colloidal quantum dots (CQD) are nanometer sized semiconductors which are dispersed in a supporting material, often a solvent, and have a tunable size-dependent energy bandgap which makes them an excellent absorber for wide band photo-absorption [10]. Although QDSC efficiency has shown a steady improvement from 3% to ~11% over the past 5 years [11], the best reported efficiency is still far below the maximum physical limit imposed by the laws of thermodynamics [12]. The demonstrated stability of QDSC is however significantly better as compared to the perovskites solar cells which are one of the major emerging PV contenders with recently demonstrated cell efficiency exceeding 20% [13]. The major contributor for the lower efficiency in QDSC has been identified as the high density ($N_t$) of sub-bandgap defect states in the CQD layer that results in poor collection efficiency and a high deficit in the open circuit voltage [7, 14]. The continuing optimizations in the cell synthesis processes, e.g., the use of novel passivation strategies [7, 15, 16], are being sought to lower $N_t$ while other techniques such as band structure engineering [17] are being investigated to implement better contact selectivity.

As the rapid process/structural innovations promise a path towards higher QDSC efficiency, there is a need to develop a thorough understanding of the physical mechanisms that influence the cell performance. In this regard, a detailed analytical model for the cell transport characteristics has been recently reported in [18] while numerical simulations have been used previously to provide guidelines for the design of high performance QDSC [19, 20]. These studies have highlighted that the best QDSC efficiency is achieved for the cells which have a relatively low carrier mobility ($\mu$) in the range of $10^{-2} - 10^{-3}$ $cm^2/Vs$ [18, 20, 21]. This behavior has been associated with the trade-off between the collection efficiency and the mobility dependent recombination, which is similar to what had been previously reported for organic solar cells [22]. An extensive modeling of this behavior for QDSC has however not been reported. The focus of this paper is to quantitatively model the role of $\mu$ for QDSC characteristics as a function of $N_t$, lowering it from the currently reported values of $\approx 10^{17} cm^{-3}$ down to what is typically achieved in bulk semiconductors ($\leq 10^{15} cm^{-3}$). In addition, we investigate energy band alignment engineering [17, 23] for QDSC to enhance efficiency beyond the limit imposed by the trade-off between mobility dependent recombination and the carrier collection. This effect is computationally explored for a broad range of design parameters such as $\mu$, $N_t$, and CQD layer thickness ($t_{CQD}$).

This paper is divided into four sections. Section II describes the modeling approach. Results are discussed in Section III and the conclusions are presented in Section IV.

## II. MODELING APPROACH

In contrast to the bulk semiconductors in which carriers flow in continuous bulk energy bands, the carrier transport in QDSC is based on the hopping mechanism between the neighboring dots [8]. For the device modeling, we use a commonly used macroscopic approach which encapsulates the microscopic details of carrier hopping into an average carrier mobility ($\mu$) that is related to the average inter-dot hopping time ($\tau_{hop}$) by [21]:

$$\mu = \frac{d^2}{6\tau_{hop}(\frac{kT}{q})} \quad (1)$$

where $d$ is the inter-dot distance and $kT$ is the average thermal energy of the carrier. In this so called 'effective medium' approach, CQD layer is treated as a 'bulk' like homogenous material so that an average dielectric constant, bandgap, and, recombination lifetime, etc., could be established. The trap-assisted recombination lifetime in this approach is given by [20]:

$$\tau_{trap} = (\sigma N_t \frac{d}{\tau_{hop}})^{-1} \quad (2)$$

where $\sigma$ is the capture cross-section, and $\frac{d}{\tau_{hop}}$ is the average thermal velocity of carriers in the CQD layer. Since $\mu$ and $\tau_{trap}$ have an opposite dependence on $\tau_{hop}$, the product ($\mu \times \tau_{trap}$) is constant and the carrier diffusion length ($L_D = \sqrt{(\frac{kT}{q})\mu\tau_{eff}}$) becomes invariant to the changes in $\mu$ when the effective carrier recombination lifetime ($\tau_{eff}$) given as: ($\tau_{eff}^{-1} = \tau_{rad}^{-1} + \tau_{trap}^{-1}$) is dominated by $\tau_{trap}$.

Incorporating the mobility dependent trap-assisted recombination, we model the QDSC characteristics using self-consistent modeling of cell electrostatics and the carrier continuity equations [24]. Simulations are done in a one dimensional simulation grid using ADEPT2.1 simulation toolkit [25]. Appendix A provides the mathematical details of the coupled electrostatic/continuity model. The QDSC structure is shown in figure 1(a). The cross-sectional stack consists of ITO/CQD-layer/buffer-layer/metal. An energy band diagram along the transport direction in a QDSC is shown in figure 1(b). The CQD layer is composed of $p$-type quantum dots with doping of $10^{16}$ cm$^{-3}$ and a bandgap of 1.2 eV. The buffer layer provides the energy band offset at CQD/metal interface as shown in figure 1(b). Such band offset in CQD layer could be physically processed through a modification in CQD ligand treatment as reported in [17]. The built-in voltage is established due to the difference in the workfunction of metal and ITO. The CQD layer thickness is kept variable to study the effect of absorber's thickness. The spectral profile of absorption coefficient for CQD layer is taken from the experimentally measured values for PbS CQD reported in [20]. The carrier mobility and recombination lifetime for electrons and holes are kept identical for simplicity. Table I shows the list the values of material parameters used in the simulation.

III.       RESULTS AND DISCUSSION

Figure 2(a) shows the dependence of short circuit current density ($J_{sc}$) on $\mu$ for a range of $N_t$ with $t_{CQD}$ = 200nm. Under short circuit condition, the whole CQD layer is depleted (as could be noted in the energy band diagram in figure 1(b)), which implies that a strong carrier drift under high electric field would dominate the carrier transport. The carrier drift length ($L_{drift} = \mu \times \varepsilon \times \tau_{eff}$) vs. $\mu$ in the middle of the CQD layer is also plotted in figure 2(a). A linear dependence of $L_{drift}$ on $\mu$ implies that the radiative recombination is dominant over the trap-assisted recombination, i.e., $\tau_{eff} \approx \tau_{rad}$. The invariance of $L_{drift}$ to $\mu$, on the other hand, implies $\tau_{eff} \approx \tau_{trap}$. Figure 2(b) elaborates the behavior of $\tau_{eff}$ vs. $\mu$. The mobility dependent regime for $\tau_{eff}$ (i.e. $\tau_{eff} \approx \tau_{trap}$) could be identified at $\mu$ greater than ~$10^{-1}$ $cm^2/Vs$ for $N_t = 10^{14} cm^{-3}$. Increasing $N_t$ to $10^{18} cm^{-3}$ results in the mobility dependence for $\tau_{eff}$ throughout the given span of mobility. As shown in figure 2(a), $L_{drift} \gg (t_{CQD} = 200nm)$ much before the start of the mobility invariant regime for $N_t = 10^{14} cm^{-3}$. For this case, $J_{sc}$ saturates to its maximum limit imposed by the net photogenerated charge collection at $\mu = 10^{-2}$ $cm^2/Vs$. For $N_t = 10^{18} cm^{-3}$, $L_{drift}$ remains smaller than $t_{CQD}$ throughout the given range of $\mu$. The maximum $J_{sc}$ for this case is therefore strongly degraded implying that a significant part of the photogenerated carriers are lost through trap-assisted recombination for the entire range of $\mu$.

The characteristics of QDSC in dark, i.e., the reverse saturation current density ($J_0$) and the ideality factor ($n$) are shown in Figure 3 as a function of $\mu$ for $N_t = 10^{14} cm^{-3}$ and $N_t = 10^{18} cm^{-3}$. The dependence of $J_0$ on $\mu$ and $N_t$ is almost linear as observed in figure 3(a) which matches to that reported in the analytical model of [18]. Moreover, $J_0$ does not show any dependence on $\Delta E$ which implies that the reverse bias transport in the dark is dominated by the minority carrier diffusion in the CQD layer and is not affected by the interfacial barrier. The ideality factor, on the other hand, shows dependence on both $N_t$ and $\Delta E$ in Figure 3(b). It is well known that $n$ is close to 1 when the carrier transport is dominated by the radiative recombination, and is close to 2 for the case when the transport is dominated by the trap-assisted recombination [26]. The values for $n$ in figure 3(b) are ~1.2 and ~2 for $N_t = 10^{14} cm^{-3}$ and $N_t = 10^{18} cm^{-3}$, respectively, and correspond to the radiative and non-radiative regimes of dominance. The presence of an energy barrier shows an effect on $n$ which is more prominent for the lower trap density ($N_t = 10^{14} cm^{-3}$). It has been reported that the presence of an interfacial barrier could result in an accumulation of carriers near the barrier under forward bias which could enhance the trap-assisted recombination and could therefore increase the ideality factor [27]. It is remarkable to note that for lower mobility ($\mu < 10^{-2}$ $cm^2/Vs$), $n$ for the case of $\Delta E = 0.2eV$ gradually decreases with lowering µ until it merges with the value of $n$ for the case of $\Delta E = 0$ eV. This phenomenon could be understood keeping in view the two contributing mechanisms of carrier transport under forward bias in the dark condition. Carriers injected from the ITO contact have to diffuse through the CQD layer until they reach the interface of CQD/buffer-layer where they undergo thermionic emission if there is an energy barrier [26]. For a high carrier mobility, carrier diffusion in CQD layer is fast and the thermionic emission over the barrier is the limiting process for the dark current. In this case, rapidly diffusing carriers in CQD accumulate near the energy barrier resulting in higher trap-assisted recombination and a consequent increase in the ideality factor as explained before [28]. As carrier mobility is decreased, slower carrier diffusion in CQD layer becomes the limiting mechanism, and for this case, there is no carrier accumulation near the interface of the energy barrier. For this case, $n$ remains identical to its value for $\Delta E = 0$.

The dependence of $V_{oc}$ on µ is shown in Figure 4 for $N_t = 10^{14} cm^{-3}$ and $N_t = 10^{18} cm^{-3}$ with and without the interfacial energy barrier. Simulation results are compared with the calculation which is based on the principle of superposition for the photocurrent and dark current under illumination [29]:

$$Voc = \frac{nkT}{q}(log \frac{J_{sc}}{J_0} + 1) \qquad (3)$$

For all simulated cases, the maximum $V_{oc}$ is observed at the lowest µ. This is because of the reason that $J_0$ is suppressed when µ is decreased as observed in figure 3(a) which increases $V_{oc}$ as given by (3). For $N_t = 10^{18} cm^{-3}$, $V_{oc}$ monotonically decreases as a function of increasing $log(µ)$, both, with and without the presence of an interfacial energy barrier. For $N_t = 10^{14} cm^{-3}$ and $\Delta E = 0$, the maximum $V_{oc}$ shows a spread over a range of µ starting from the lowest mobility ($10^{-4} cm^2/Vs$) up to $µ \leq 10^{-2} cm^2/Vs$. This spread is related to the increase in $J_{sc}$ as µ is increased in this range (see figures 2(a)). When the interfacial energy barrier is present ($\Delta E = 0.2\ eV$) for $N_t = 10^{14} cm^{-3}$, $V_{oc}$ peak spreads over a wider range of µ which extends from $10^{-4} cm^2/Vs$ up to $µ \approx 10^{-1} cm^2/Vs$. This behavior is qualitatively consistent with the increase in both $J_{sc}$ and $n$ as a function of increasing µ for $N_t = 10^{14} cm^{-3}$ in the presence of the energy barrier (see figure 3(b)). Physically, it corresponds to the suppression of dark current due to the energy barrier under forward bias. It should be noted that the trend of the calculated $V_{oc}$ vs. µ qualitatively matches with that of the simulated $V_{oc}$ except for the case of $N_t = 10^{14} cm^{-3}$ with $\Delta E = 0.2\ eV$. Quantitatively, the difference between simulated and calculated $V_{oc}$ is expected since the superposition of the IV curve in the dark and under illumination, which forms the basis of (3), could fail to apply for thin film solar cells [30, 31]. It is well-known that the photocurrent in thin films solar cells could show voltage dependence below the open circuit voltage, and the injected current from the contacts may not be identical under dark and illumination conditions [32]. Figure 4(b) indicates that the deviation of the behavior between the simulated and calculated $V_{oc}$ is most significant for the case of lower trap density in the presence of interfacial energy barrier.

Figure 5 shows the contour plots of $V_{oc}$, $J_{sc}$, $FF$, and, $\eta$ as a function of $N_t$ and $\mu$ for $\Delta E = 0 \ eV$ and $0.2 \ eV$. Results for two different CQD thicknesses ($t_{CQD} = 200 nm$, and $t_{CQD} = 1 \mu m$) are shown for comparison. The behavior for $V_{oc}$ and $J_{sc}$ corresponds to the earlier explanations with regards to figures 2 – 4. At lower $\mu$, $FF$ shows an improvement with increasing $\mu$ which is similar to the behavior of $J_{sc}$. This is because $J_{sc}$ and $FF$ both depend on the carrier diffusion length in CQD layer which improves with increasing $\mu$ and lowering $N_t$. At higher $\mu$, the increase of $J_{sc}$ saturates while $FF$ shows a slight degradation similar to that observed for $V_{oc}$. For $N_t < 10^{16} cm^{-3}$, a significant increase in $J_{sc}$ can be observed for the thicker $t_{CQD}$ as compared to the thinner, due to the increased photo-absorption in the former. Owing to an increased $J_{sc}$, $\eta$ shows a significant increase for the thicker CQD cell. For $N_t > 10^{16} cm^{-3}$, the benefit of enhanced photogeneration in thicker CQD cell diminishes due to significant trap-assisted recombination loss. The optimal $\mu$ for the best cell efficiency is in the range $10^{-2} - 10^{-1} \ cm^2/Vs$ for $\Delta E = 0 \ eV$, and $10^{-1} - 10^{0} \ cm^2/Vs$ for $\Delta E = 0.2 \ eV$. At the optimal $\mu$, the best values for $J_{sc}$ and $V_{oc}$ and $FF$ coincide as observed in the contour plots of figure 5.

Figure 6 elaborates the dependence of $\eta$ on $t_{CQD}$ at an optimal $\mu$ ($10^{-1} \ cm^2/Vs$) with and without an interfacial barrier. A significant improvement in $\eta$ as a function of $t_{CQD}$ could be observed which saturates to its peak at $t_{CQD} \approx 2 \mu m$. For $t_{CQD} < 1 \mu m$, presence of an interfacial barrier ($\Delta E = 0.2 \ eV$) results in a significant improvement for $\eta$ as compared to the case when there is no interfacial barrier. At $t_{CQD} = 200 \ nm$, the improvement in $\eta$ due to the interfacial barrier is ~37%. The effect of interfacial barrier on $\eta$ gradually diminishes as $t_{CQD}$ is increased. The inset of Figure 6 shows the comparison of the simulated results (blue circles) with the experimental data (red triangles) reported in [20]. The values for $\mu$ and $N_t$ in the simulated curve are $0.02 \ cm^2/Vs$ and $2 \times 10^{16} cm^{-3}$ respectively which are chosen to match with those reported in the experimental data [20, 33]. The simulation vs. experiment match is qualitatively reasonable given the experimental uncertainty in the values of $\mu$ and $N_t$, and, a relatively simple model assumed in our simulations in which the secondary effects such as series resistance in the cell, surface reflectance, and parasitic absorbance in the window layers are ignored.

## IV. CONCLUSIONS

In summary, we show that the physical design of colloidal quantum dot solar cells for enhanced efficiency requires a careful optimization of the carrier mobility. The optimal mobility could modulate with CQD layer thickness and the interfacial energy band alignment within the range of $10^{-2} \, cm^2/Vs - 10^0 \, cm^2/Vs$. An interfacial energy band alignment at the metal/CQD contact could enhance the open circuit voltage by reducing the dark current under forward bias. This effect is prominent in the regime of low ($< 10^{16} cm^{-3}$) trap density for thinner ($< 1 \mu m$) CQD layer and intensifies as CQD layer thickness is decreased. For thicker CQD thickness ($> 1 \mu m$), carrier diffusion in CQD layer becomes the transport bottleneck which diminishes the effect of interfacial band alignment on open circuit voltage. These considerations, in particular, become important as continuous improvements in the cell synthesis aspire to reduce the trap density in the CQD layer from today's experimental values of $\sim 10^{17} cm^{-3}$ to values below $10^{16} cm^{-3}$. The physical insights provided in this work could be useful for the future development of the quantum dot solar cell technology.

### Table I. SIMULATION PARAMETERS

| Parameter | Value |
|---|---|
| Effective Bandgap (CQD) | $1.2 \, eV$ |
| Dielectric constant (CQD) | 11.7 |
| Electron affinity (CQD) | $4.05 \, eV$ |
| Effective density of states $Nc$, $Nv$ (CQD) | $1 \times 10^{20} \, cm^{-3}$ |
| CQD doping ($p$-type) | $1 \times 10^{16} \, cm^{-3}$ |
| Inter-dot distance | 7nm |
| Capture cross section (SRH recombination) | $1 \times 10^{13} \, cm^2$ |
| Radiative recombination coefficient (CQD) | $1 \times 10^{-10} \, cm^{-3} s^{-1}$ |
| Workfunction ITO [34] | $3.8 \, eV$ |
| Workfunction metal | $4.95 \, eV$ |

### APPENDIX

A. Simulation details

The Poisson equation is given by [24]:

$$\nabla^2 V(x) = \frac{q}{\epsilon_{Si}}[N_D - N_A + p(x) - n(x)] \quad (A1)$$

where $V$ is the electrostatic potential, $q$ is the charge on electron, $N_D$ ($N_A$) are donor (acceptor) doping density, $\epsilon_{Si}$ is the permittivity of silicon, $n$ and $p$ are position dependent electron's and hole's concentration respectively. The steady state continuity equation for electrons (holes) under drift-diffusion formalism are given by:

$$D\frac{\partial^2 n}{\partial x^2} + \mu\varepsilon(x)\frac{\partial n}{\partial x} + G(x) - R(x) = 0 \quad (A2)$$

$$D\frac{\partial^2 p}{\partial x^2} - \mu\varepsilon(x)\frac{\partial p}{\partial x} + G(x) - R(x) = 0 \quad (A3)$$

where $\varepsilon$ is the electric field, $G$ and $R$ are carrier generation and recombination rates, and, $\mu$ and $D$ are the carrier mobility and diffusion constant respectively.

## ACKNOWLEDGEMENT

This work is based in part on work supported by Faculty Initiative Fund (FIF) award, made by University Research Council of Lahore University of Management Science (LUMS).

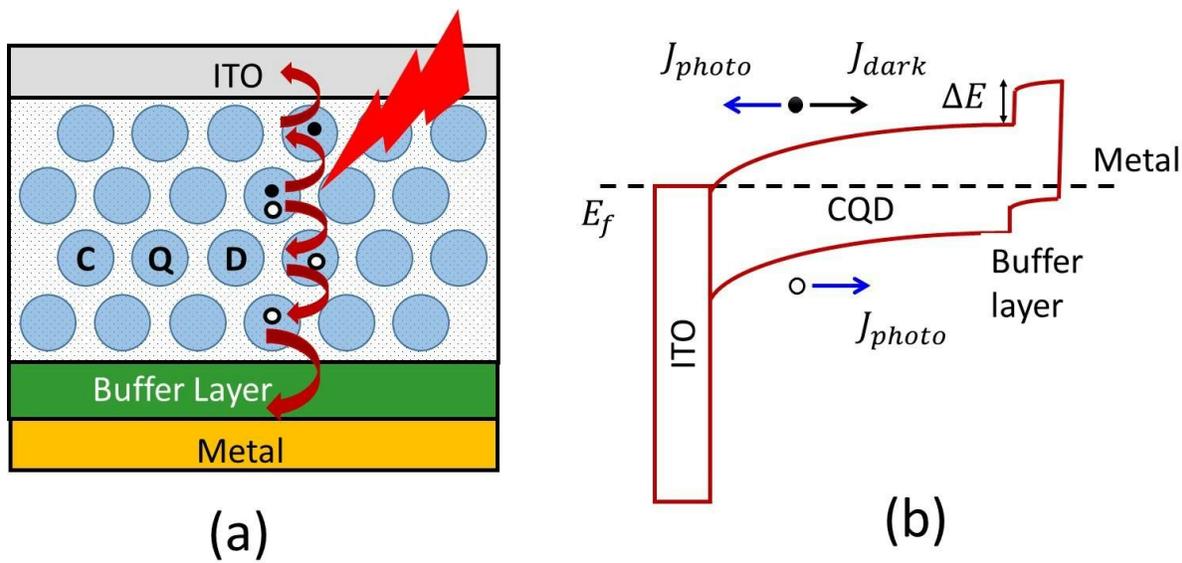

Fig. 1.  (a) Schematic of device architecture and hopping mechanism of photogenerated carriers to their respective contacts. (b) Illustration of bulk like energy band diagram of quantum dot solar cell using an effective medium approach. Minority carrier diffusion in CQD describes the transport mechanism under dark. A buffer layer represents a thin CQD layer having an energy band offset with the bulk CQD absorber. Under short circuit condition, depletion extends throughout the thickness of CQD layer. $\Delta E$ is the barrier height that curtails dark current under forward bias hence increasing the open circuit voltage.

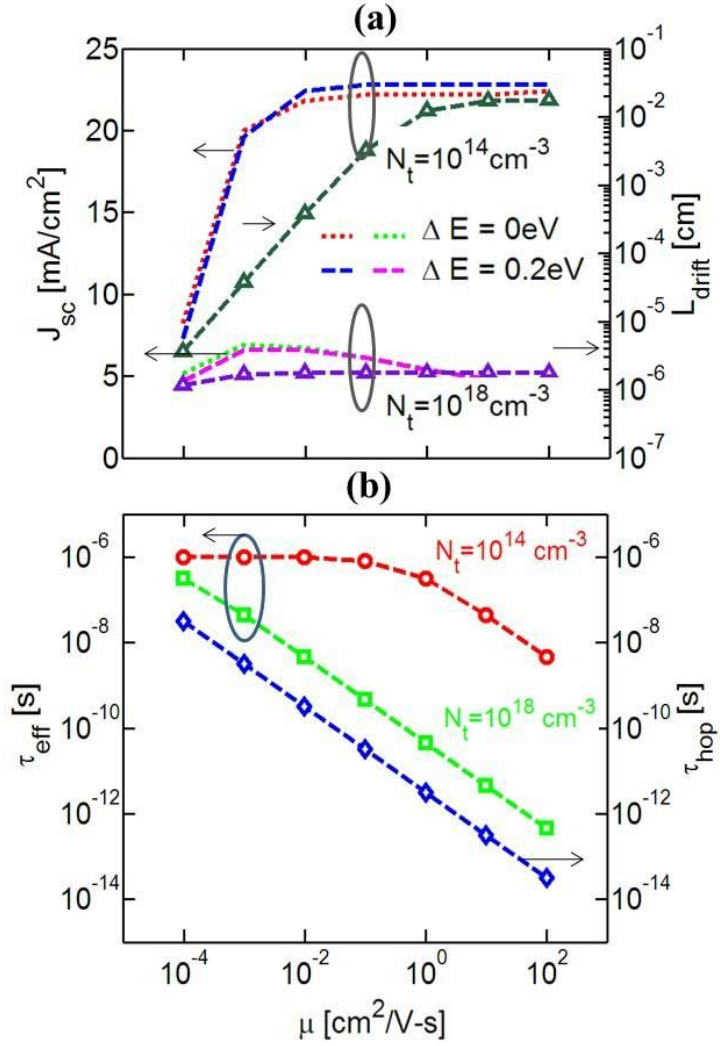

Fig. 2. (a) Dependence of $J_{sc}$ and $L_{drift}$ on $\mu$ for a range of $N_t$ and $t_{CQD}=200nm$. With increasing $\mu$ and for a fewer traps, carriers drift longer distance, improving $J_{sc}$ with it. For higher trap densities, $J_{sc}$ is significantly lower, owing to a strong carrier loss through trap-assisted recombination. At higher $\mu$, $J_{sc}$ saturates to its peak defined by the maximum collection of the photo-absorption flux in the CQD layer. An energy band offset ($\Delta E=0.2eV$) at CQD/metal interface improves the saturated $J_{sc}$ for the case of lower trap density by reducing electron loss into the metal contact. (b) Relationship between $\mu$ and $\tau_{eff}$ is subject to the dominant mode of recombination. While $\mu$ is directly proportional to $\tau_{hop}$, an inverse proportionality exists between trap assisted recombination lifetime ($\tau_{trap}$) and $\tau_{hop}$. With increasing $N_t$, $\tau_{trap}$ dominates and the product $\mu \times \tau_{eff}$ therefore stays constant as long as ($\tau_{eff} \approx \tau_{trap}$).

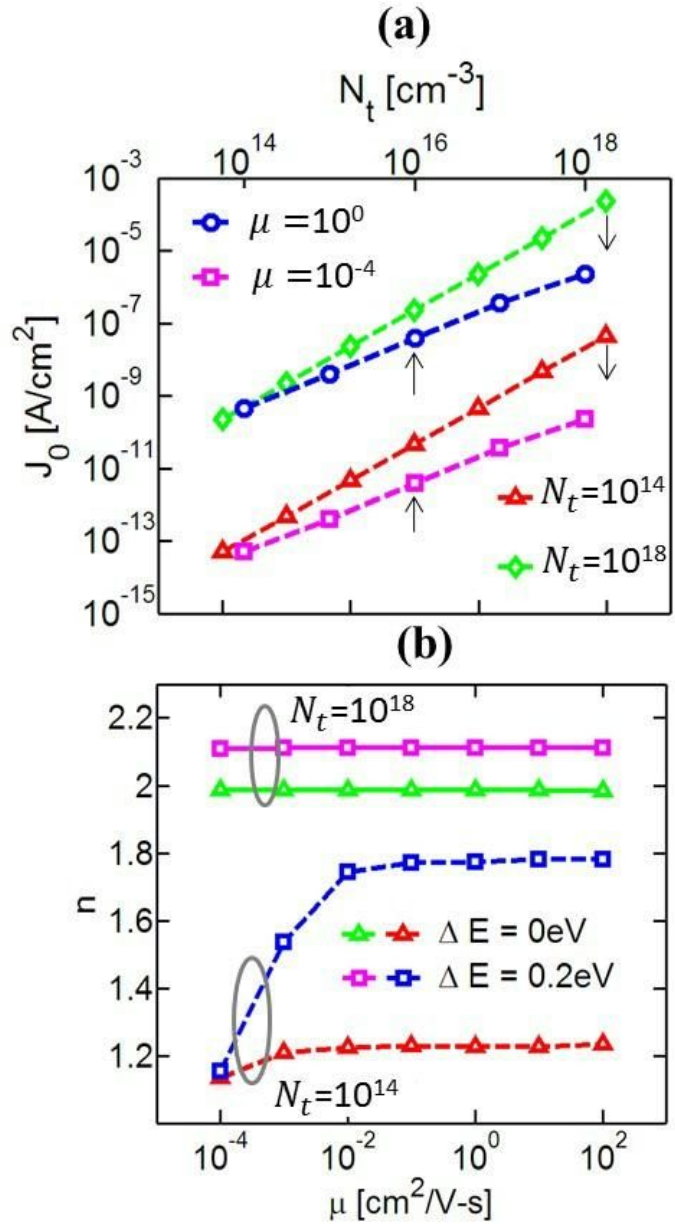

Fig. 3. (a) Relationship of $J_0$ with $\mu$ and $N_t$ varies linearly irrespective of $\Delta E$ confirming that the reverse bias dark current is limited by minority carrier diffusion. (b) The ideality factor ($n$) vs. $\mu$ is however affected by $\Delta E$. The presence of the barrier ($\Delta E = 0.2 eV$) results in carrier accumulation near the barrier which enhances trap-assisted recombination and hence increases $n$. This effect is more prominent at lower $N_t$. At very low $\mu$ and for lower $N_t$, the forward bias dark current becomes diffusion limited and is unaffected by the presence of barrier which makes $n$ identical for the two cases of $\Delta E$.

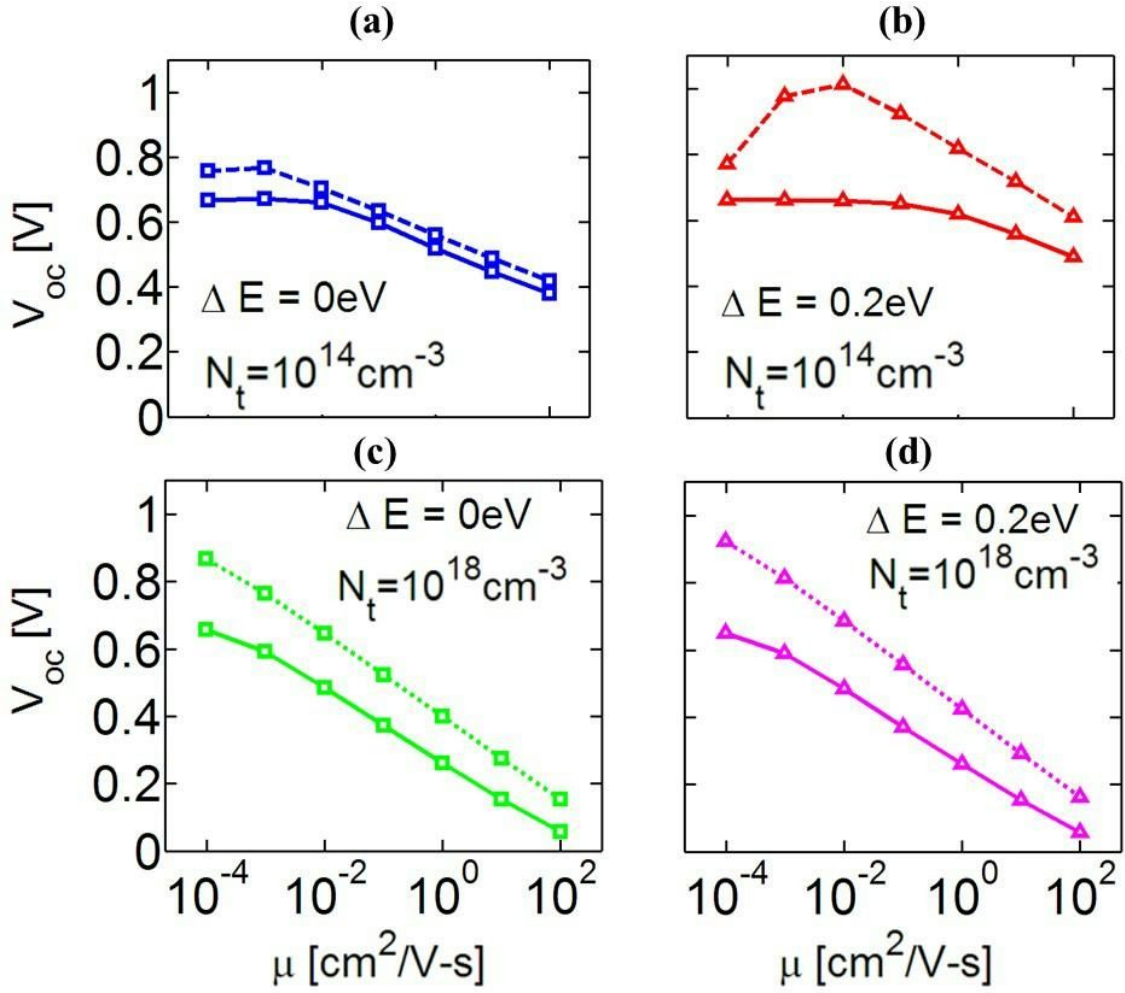

Fig. 4. (a) Relationship of $V_{oc}$ and $\mu$ for (a) $N_t = 10^{14} cm^{-3}$ and $\Delta E = 0\ eV$, (b) $N_t = 10^{14} cm^{-3}$ and $\Delta E = 0.2\ eV$, (c) $N_t = 10^{18} cm^{-3}$ and $\Delta E = 0\ eV$, and, (d) $N_t = 10^{18} cm^{-3}$ and $\Delta E = 0.2\ eV$. The solid lines are simulation results while the dashed lines are calculated using the superposition of dark and light-generated current as given by (3). The qualitative match between simulation and calculation is good for (a), (c), and (d). The quantitative difference implies that the superposition principle fails to apply well for thin film QDSC.

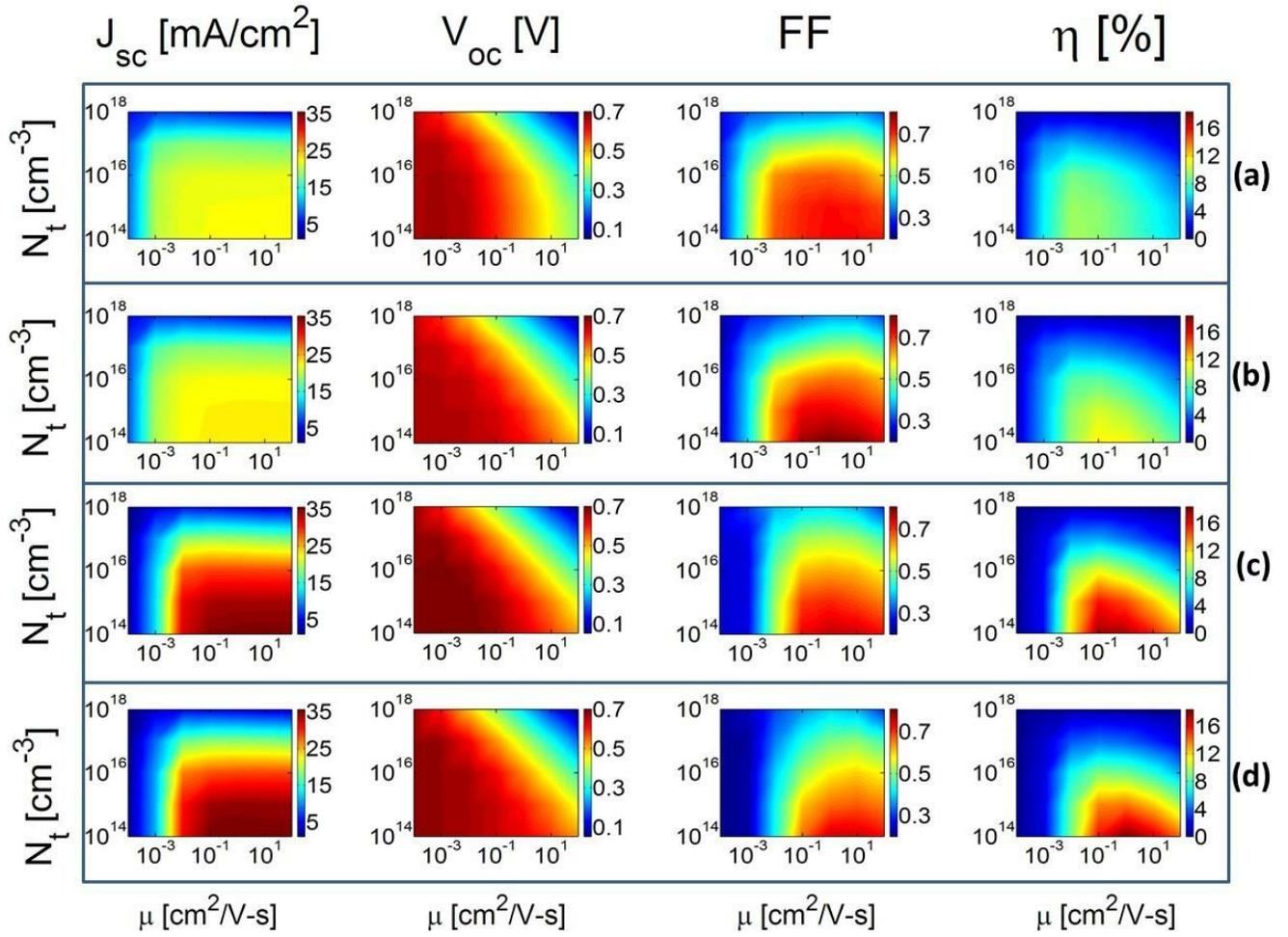

Fig. 5. Contour plots of $J_{sc}$, $V_{oc}$, $FF$, and, $\eta$ as a function of $N_t$ and $\mu$: (a) For $t_{CQD}=200nm$ and $\Delta E=0eV$, a maximum efficiency of ≈9% is achieved. (b) For $t_{CQD}=200nm$ and $\Delta E=0.2eV$, $J_{sc}$, $V_{oc}$, and, $FF$ improve to give a relatively higher efficiency of 11.4%. For (c) and (d), $t_{CQD}$ is increased to 1$\mu m$ to allow for an increased photo-absorption. In (c), overall efficiency of the device for $\Delta E=0eV$ improves to η≈17%. With an interfacial barrier ($\Delta E=0.2eV$) in (d), wrong contact carrier extraction is curtailed to give an enhanced efficiency of about 18%.

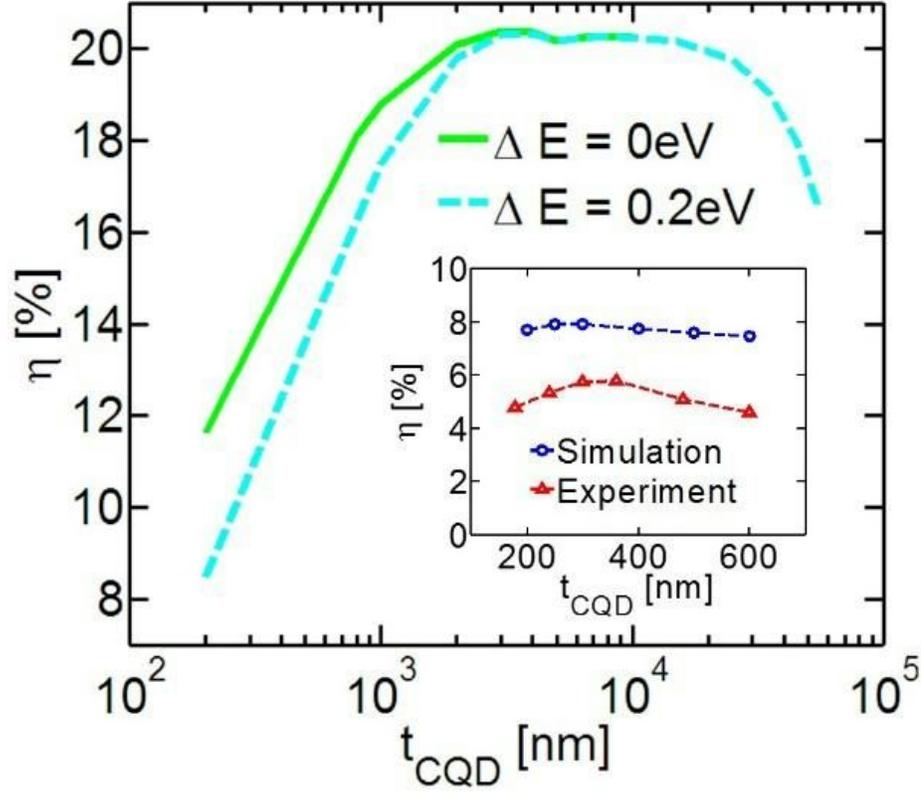

Fig. 6. Dependence of $\eta$ on $t_{CQD}$ for two cases of $\Delta E$ at an optimal $\mu$ ($10^{-1}$ $cm^2/Vs$) and $N_t = 10^{14} cm^{-3}$. Under these conditions, an increase in photo-absorption with increasing device thickness results in higher $J_{sc}$ and improved $\eta$. The interfacial barrier ($\Delta E = 0.2$eV) suppresses dark current for thinner ($t_{CQD} \leq 1\mu m$) CQD and improves $\eta$ due to an enhanced open circuit voltage. For $t_{CQD} \gg 1\mu m$, the effect of barrier on $\eta$ is negligible since the forward bias dark current becomes diffusion limited for thicker CQD and becomes independent of the barrier. The inset in the figure shows the behavioral match of simulation results and experimental data [20]. The simulation assumes $\mu = 0.02$ $cm^2/Vs$ and $N_t = 2 \times 10^{16} cm^{-3}$ which are similar to those reported for the experiment. The behavioral match between simulation and experiment is reasonable given that the simulation neglects secondary effects such as series resistance, surface reflectance, and parasitic absorption in the window layers.